\documentstyle[prb,aps,twocolumn]{revtex}
\begin{document}
\draft
\twocolumn[
\hsize\textwidth\columnwidth\hsize\csname @twocolumnfalse\endcsname
\title{Destruction of Peierls dimerization in the molecular crystal
model:\\ Effects of quantum phonon fluctuations}
\author{C.Q. Wu,$^*$ Q.F. Huang, and X. Sun}
\address{T.D. Lee Physics Laboratory and Department of Physics,\\
Fudan University, Shanghai 200433, China}
\maketitle
\centerline{(Received 2 July 1995)}
\begin{abstract}

Effects of quantum phonon fluctuations on the Peierls dimerization in
the
one-dimensional molecular crystal model are reexamined by a
functional integral
approach. An equation for the dimerization order parameter
is obtained within a one-loop approximation. The destruction of
Peierls
dimerization is found by taking the quantum phonon fluctuations into
account for the spin-1/2 electron system. The consistency and
discrepancy of
the result with previous ones are discussed.

\end{abstract}

]
In a one-dimensional electron-phonon system, the metallic phase is
unstable
against the electron-phonon interaction, which results in the Peierls
transition.\cite{pei} At the half-filling case, a dimerized lattce
will be
formed and the ground state is of a dimerized long-range order. This
is
true for an arbitrarily small electron-phonon coupling within the
mean-field
adiabatic approximation which treats the phonon degree of freedom
classically.
It is an interesting problem whether the dimerized ground state
survives
the quantum phonon fluctuations. There have been many works to study
this problem. Among them are the renormalization group analysis
plus the Monte Carlo numerical calculations\cite{fh,hf} on the
Su-Schrieffer-Heeger (SSH) model\cite{ssh} and the molecular crystal
(MC)
model,\cite{mc} the variational approach\cite{zhe,zheng} on the same
models, and
the one-loop approximation on the SSH model\cite{sch} and the halogen-
bridged
mixed-valence linear chain complexes (MX) model.\cite{huang}

In the SSH model as well as the MX model, the optical phonon-electron
interactions are dominant and the acoustic phonon-electron
interactions
are weak, so the quantum fluctuations are mainly induced by the
optical
phonon-electron couplings. It has been shown that the dimerization of
lattice
survives the quantum phonon fluctuations for the spin-$\case{1}/{2}$
systems whereas
it is broken for the spinless systems.\cite{fh,huang} However, in the
MC model,
the interactions between electrons and two branches of phonons are
equally
important, it seems that the quantum fluctuations should be much
stronger than
that in the SSH and MX models. Although the renormalization-group
analysis
shows absence of dimerized long range order for the zero mass system,
Hirsch
and Fradkin concluded that the ground state is dimerized for any
nonzero mass
systems from the analysis of symmetries as well as their Monte Carlo
numerical
simulations.\cite{hf}

In this paper, we reexamine the effects of quantum phonon
fluctuations on
Peierls dimerizaton in the molecular-crystal model by a functional
integral
approach. An equation for the dimerization order parameter is
obtained within
a one-loop approximation. We find the Peierls dimerization will be
destroyed
by taking the quantum phonon fluctuations into account for the
spin-$\case{1}/{2}$ electron system. An equivalency between the
spin-$\case{1}/{2}$ MC model and the spinless SSH or MX model is
shown. The
consistency and discrepancy of our results with the renormalization-
group
analysis as well as the Monte Carlo numerical simulations are
discussed.

The one-dimensional molecular-crystal model we consider in this paper
takes
the following form,\cite{mc}
\begin{eqnarray}
H=&&\sum_l \left [\frac{1}{2M}p_l^2+\frac{1}{2}Ku_l^2 \right
]\nonumber\\
&&-\sum_{l,s}t_0(c^\dagger_{l,s}c_{l+1,s}+c^\dagger_{l+1,s}c_{l,s})
+\sum_{l,s}\alpha u_l c^\dagger_{l,s}c_{l,s}.
\end{eqnarray}
where $c^\dagger_{l,s}$ and $c_{l,s}$ are the creation and
annihilation
operators of electrons sitting at molecular $l$ with spin $s$, $u_l$
(its
conjugated momentum is $p_l$) describes an internal degree of freedom
of molecule $l$, $t_0$, $\alpha$ and $K$ are the constants for
the electron hopping, electron-phonon coupling and lattice elasticity,
and $M$ is the mass of the internal vibrations in molecules which
form the
one-dimensional lattice.
The model can also be viewed as describing the coupling between
electrons
and excitons with boson characters.

To take into account the quantum phonon fluctuations, we employ a
functional
integral approach. The partition function of the system can be
written as a
functional integral over both complex and Grassmann variables:
\begin{equation}
Z=\int {\cal D} [u]{\cal D} [\phi^*\phi] e^{-S},
\end{equation}
where the action $S$ is
\begin{eqnarray}
S&&=\int_0^\beta d\tau \{ \sum_l
[\case{1}/{2}K u_l^2+\case{1}/{2}M(\partial
u_l/\partial\tau)^2]\nonumber\\
&&+\sum_{l,s}\phi_{l,s}^*(\partial_\tau-\mu)\phi_{l,s}\nonumber\\
&&-
\sum_{l,s}t_0(\phi_{l,s}^*\phi_{l+1,s}+\phi_{l+1,s}^*\phi_{l,s})
\nonumber\\
&&+\sum_{l,s}\alpha u_{l}\phi_{l,s}^*\phi_{l,s}\}.
\end{eqnarray}
By making Fourier transformations on the integral variables
\begin{equation}
u_l(\tau)=\frac{1}{\sqrt{\beta L}}\sum_{q,\Omega}
e^{-i(ql+\Omega\tau)}u_q(\Omega),
\end{equation}
and similarly for $\phi^*_{l,s}$ and $\phi_{l,s}$, then
for the phonon variables we have two branches on a dimerized lattice,
that is, the acoustic phonon $a_q(\Omega) [\equiv u_q(\Omega)]$ and
the
optical phonon $b_q(\Omega) [\equiv u_{q+Q}(\Omega)]$, $Q=2k_F$
($k_F=\pi/2$ for the half-filled band case, the lattice constant has
been set as unit). For the Grassmann variables we could introduce a
spinor notation via\cite{swy}
\begin{equation}
\Phi_{k,s}=\left(\begin{array}{c} \phi_{k+k_F,s}\\
                        \phi_{k-k_F,s}\end{array}\right),
\end{equation}
which are right- and left-moving electrons, respectively,
then we have the partition function as
\begin{equation}
Z=\int {\cal D} [a^*a,b^*b,\Phi^*\Phi] e^{-S},
\end{equation}
and the corresponding action could be written as a sum of five terms
\begin{equation}
S=S_a+S_b+S_e+S_{a-e}+S_{b-e},
\end{equation}
where the acoustic- and optical-phonon parts of the action are,
respectively,
\begin{mathletters}
\begin{eqnarray}
S_a&=&\sum_{q,\Omega}\left [\case{1}/{2}K+\case{1}/{2}M\Omega^2\right
]a_q(\Omega)a_{-q}(-\Omega),\\
S_b&=&\sum_{q,\Omega}\left [\case{1}/{2}K+\case{1}/{2}M\Omega^2\right
]b_q(\Omega)b_{-q}(-\Omega),
\end{eqnarray}
the electronic part of the action is
\begin{equation}
S_e=\sum_{k,\omega,s}\Phi^*_{k,s}(\omega)[(i\omega-\mu)+2t_0\sin k
\sigma_3]
\Phi_{k,s}(\omega),
\end{equation}
and the acoustic and optical phonon-electron interaction parts are
the following
\begin{eqnarray}
S_{a-e}&=&\frac{\alpha}{\sqrt{\beta
L}}\sum_{q,\Omega}\sum_{k,\omega,s}
a_q(\Omega)\Phi^\dagger_{k,s}(\omega)\Phi_{k-q,s}(\omega-\Omega),\\
S_{b-e}&=&\frac{\alpha}{\sqrt{\beta
L}}\sum_{q,\Omega}\sum_{k,\omega,s}
b_q(\Omega)\Phi^\dagger_{k,s}(\omega)\sigma_1\Phi_{k-q,s}(\omega-
\Omega),
\end{eqnarray}
\end{mathletters}
where $\sigma_i$ are Pauli matrices.

In the SSH model and the MX model, the optical phonon-electron
interactions
are dominant and the acoustic phonon-electron interactions can be
ignored.\cite{huang,swy} However, as seen from the Eq. (8d) and (8e),
the
couplings between electrons and two branches of phonons are
equivalent in
the MC model. The fact implies we should take the both into our
consideration.
One can integrate out the electronic variables of
the partition since the action is bilinear in the Grassmann fields,
the
resulting partition function is
\begin{equation}
Z=\int {\cal D} [a^*ab^*b] e^{-S_{\rm eff}},
\end{equation}
and the effective action is
\begin{eqnarray}
S_{\rm eff}=\sum_{q,\Omega}&&\left
[\case{1}/{2}K+\case{1}/{2}M\Omega^2\right ]
\left [a_q(\Omega)a_{-q}(-\Omega)+b_q(\Omega)b_{-q}(-\Omega)\right]
\nonumber\\
&&-N\ln\det(\cal{M}),
\end{eqnarray}
where $N$ is the spin degree of freedom, $\det$ denotes the
determinant of
the matrix $\cal{M}$, which is defined by the actions $S_e+S_{a-
e}+S_{b-e}=
\sum_s\Phi_s^\dagger\cal{M}$$\Phi_s$ and has $k$, $\omega$, and the
indices
of the Pauli matrices as labels.

By the functional derivation of the effective action $S_{\rm eff}$
with
the phonon variables vanishing, we have the Peierls dimerization,
i.e., the
optical phonon condensation at zero momentum and zero frequency
$\langle
\alpha b_q(\Omega)\rangle=\sqrt{\beta
L}\Delta\delta_{q,0}\delta_{\Omega,0}$,
where the phonon order parameter $\Delta$ describes the Peierls
dimerization.
Defining the noninteracting electronic Green's function
$G_0(k,\omega)$ by
\begin{equation}
G_0(k,\omega)=-(i\omega-\mu+2t_0\sin k \sigma_3+\Delta\sigma_1)^{-1},
\end{equation}
and $b_q(\Omega)=\langle b_q(\Omega)\rangle+\tilde{b}_q(\Omega)$,
we obtain the effective action by expanding the logarithm in order of
$\tilde{b}$ and $a$:
\begin{equation}
S_{\rm eff}=S_{\rm eff}^{(0)}+\sum_{n=1}^{\infty}S_{\rm eff}^{(n)}.
\end{equation}
$S_{\rm eff}^{(0)}$ is the zeroth-order contribution in $\tilde{b}$
and $a$ and
is given by
\begin{equation}
S_{\rm eff}^{(0)}=N\beta L
\left\{\frac{\Delta^2}{2\lambda\pi t_0}-
\int\frac{dkd\omega}{(2\pi)^2}\ln\det
G_0^{-1}(k,\omega)\right\},
\end{equation}
in the thermodynamic limit, the dimensionless electron-phonon coupling
constant $\lambda$ is defined by $\lambda=N\alpha^2/\pi Kt_0$, the
chemical
potential $\mu=0$ at the half-filling case. The electronic spectrum
$E_k=\pm\sqrt{(2t_0\sin k)^2+\Delta^2}$, the electronic gap is
$2\Delta$.

The mean field gap equation
\begin{equation}
\Delta_{\rm ad}=\Lambda e^{-2/\lambda}
\end{equation}
follows immediately from $\partial S_{\rm
eff}^{(0)}/\partial\Delta=0$,
where $\Lambda$ is the integral cutoff. The mean-field result tell us
that the
ground state is dimerized for any electron-phonon interactions. To
investigate
the effects of quantum phonon fluctuations, we should include higher
order contributions.
The first-order term $S_{\rm eff}^{(1)}$ vanishes since $S_{\rm
eff}^{(0)}$ is
obtained from the saddle point approximation. The second-order
contribution
to the effective phonon action is
\begin{eqnarray}
S_{\rm eff}^{(2)}=&&\sum_{q,\Omega}\left\{a^*_q(\Omega)a_{q}(\Omega)
\left[\case{1}/{2}M\Omega^2+\lambda K
f(q,\Omega)\right]\right.\nonumber\\
&&\left.+\tilde{b}^*_q(\Omega)\tilde{b}_{q}(\Omega)
\left[\case{1}/{2}M\Omega^2+\lambda K g(q,\Omega)\right]\right\},
\end{eqnarray}
where the functions $f(q,\Omega)$  and $g(q,\Omega)$ are defined as
follows
\begin{mathletters}
\begin{eqnarray}
f(q,\Omega)=\frac{1}{2\lambda}+\frac{\pi t_0}{2}
\int\frac{dkd\omega}{(2\pi)^2}{\rm
Tr}[&&G_0(k-q,\omega-\Omega)\nonumber\\
&&G_0(k,\omega)],\\
g(q,\Omega)=\frac{1}{2\lambda}+\frac{\pi t_0}{2}
\int\frac{dkd\omega}{(2\pi)^2}{\rm
Tr}[&&G_0(k-q,\omega-\Omega)\nonumber\\
&&\sigma_1G_0(k,\omega)\sigma_1].
\end{eqnarray}
\end{mathletters}
\indent By performing the integration in Eq. (9) over the
fluctuations $\tilde{b}$
and $a$ to the second-order term, we have
\begin{equation}
Z=e^{-\beta L \Gamma[\Delta]},
\end{equation}
where the free energy density $\Gamma[\Delta]$ is composed of two
parts,
\begin{equation}
\Gamma[\Delta]=\Gamma_0[\Delta]+\Gamma_1[\Delta],
\end{equation}
$\Gamma_0[\Delta]=S^{(0)}_{\rm eff}/\beta L$ is the zeroth-order
contribution, and the one-loop contribution $\Gamma_1[\Delta]$
is given by
\begin{eqnarray}
\Gamma_1[\Delta]=\frac{1}{2}\int\frac{dqd\Omega}{(2\pi)^2}&&\left\{
\ln\left[\case{1}/{2}M\Omega^2+\lambda K
f(q,\Omega)\right]\right.\nonumber\\
&&\left.+\ln\left[\case{1}/{2}M\Omega^2+\lambda K
g(q,\Omega)\right]\right\},
\end{eqnarray}
in the zero-temperature and thermodynamic limit. The equation for the
phonon order parameter $\Delta$ is determined by
$\partial\Gamma[\Delta]/\partial\Delta=0$.
Within the one-loop approximation, we have
\begin{eqnarray}
1=2\pi\lambda
t\int\frac{dqd\Omega}{(2\pi)^2}&&\left\{\frac{1}{\Omega^2+E^2_{\rm
q}}\right.\nonumber\\
&&\left.-\frac{1}{N}\cdot\frac{R_f^{-1}(q,\Omega)+R_g^{-
1}(q,\Omega)}{4\Delta}\right\},
\end{eqnarray}
where
\begin{equation}
R_u(q,\Omega)\equiv
\frac{\Omega^2}{\Omega_0^2u'(q,\Omega)}+\frac{u(q,\Omega)}
{u'(q,\Omega)},
\end{equation}
with both $u=f$ and $g$, the renormalized phonon frequency
$\Omega_0^2=2\lambda\omega_Q^2$ ($\omega_Q\equiv\sqrt{K/M}$).
In the limit of weak electron-phonon interaction, only electrons near
the
Fermi surfaces be important, that is, we can take small momentum
approximation
in the calculation of the functions $f(q,\Omega)$ and $g(q,\Omega)$,
which gives $f(q,\Omega)=g(q,\Omega)$. This result implies the
equivalence
of the contributions from the optical and acoustic phonons in quantum
fluctuations in the MC model. The derivation of the function
$u(q,\Omega)$ with the phonon order parameter $\Delta$ can be
calculated
straightforwardly, it gives
$u'(q,\Omega)/u(q,\Omega)=4\Delta/(\Omega^2+4t_0^2q^2+4\Delta^2)$.
Then we have
\begin{equation}
R_u(q,\Omega)=(d^2\Omega^2+4t_0^2q^2+4\Delta^2)/4\Delta,
\end{equation}
with the abbreviation $d^2=1+4\Delta^2/\Omega_0^2$, the higher-order
terms
have been neglected as usual (in a renormalization-group study). The
equation (20)
can be compared with that in the SSH or MX model where only one $R_u$
appears
since the acoustic phonon-electron interaction can be ignored, so that
the spin-$\case{1}/{2}$ ($N=2$) MC model is equivalent to the
spinless ($N=1$)
SSH or MX model in this sense.

Now we obtain the phonon order parameter
\begin{equation}
\Delta=\Lambda \exp\left[-2\left(1-\frac{\lambda\ln
2}{Nd}\right)\left/\lambda\left(1-
\frac{2}{Nd}\right)\right]\right.,
\end{equation}
by performing the integral of Eq. (20). It can be seen that in the
adiabatic
limit $M\rightarrow\infty$, the quantum fluctuation is completely
suppressed
and the above equation returns to the mean-field one (14).

The results obtained from Eq. (23) are discussed as follows. (1) For
the spinless electron systems,
the ground state is undimerized for a weak electron-phonon coupling
whereas
it is dimerized for a strong electron-phonon coupling. The numerical
results
are in good agreement with that of the Monte Carlo
simulations.\cite{hf} (2) For
the spin-$\case{1}/{2}$ electron systems, the ground state is
undimerized in
the limit of $M=0$; it is dimerized for the system with any nonzero
mass in the limit of strong electron-phonon interactions. This result
agrees with that of the renormalization-group analysis.\cite{hf} (3)
A typical
curve of our numerical results for the $N=2$ case is given in Fig.~1.
To
compare the data we obtain with that of the Monte Carlo calculation,
we used
the phonon-staggered order parameter $m_p$, which is related to
$\Delta$
through $m_p=\Delta/\alpha$. It is found that the Peierls
dimerization is
destroyed in weak electron-phonon coupling regime. This is consistent
with
the spinless SSH or MX model since they are equivalent. The data we
obtained
are quite close to that of Monte Carlo simulation,\cite{hf} in which
a jump
in $m_p$ is seen at a critical $\alpha$. We believe that this is a
signature
of the transition discussed in the present work, although an opsite
conclusion
was reached at by the original authors. This result also implies that
the break
of the continuous symmetry in spin space does not necessarily produce
dimerization
although the continuous symmetry does prevent the existence of long-
range
dimerization order in the model.\cite{hf} Finally, we mention the
works by a variational approach,\cite{zhe,zheng} which can not give
out a
transition since it underestimates the quamtun fluctuations.

In summary, the effects of quantum phonon fluctuations on Peierls
dimerization
in the molecular crystal model are reexamined by a functional
integral approach.
The calculation is performed by first integrating out the electronic
variables
of the partition function and then expanding the effective action to
the
quadratic terms in the phonon variables. It is found that the Peierls
dimerization is destroyed by the quantum phonon fluctuations for the
spin-$\case{1}/{2}$ electron systems. This result
favors earlier suggestion by Little\cite{litt} on the possible high-
temperature
superconductivity in this kind of systems.

The authors are grateful to the Shanghai Qi-Ming-Xing Plan for Young
Scientists, the State Education Commission of China, and the National
Nature Science Foundation of China for their support.

\begin{figure}
\caption{The dependence of the phonon-staggered order parameter $m_p$
on
the electron-phonon coupling $\alpha$ for $\omega_Q=1.1$, $K=0.25$,
and $t_0=1$. The dashed line is the mean-field result and the data
with error
bar are the Monte Carlo results.}
\end{figure}

\end{document}